\documentclass[12pt,a4paper]{article}

\newcommand{\be}{\begin{equation}}
\newcommand{\ee}{\end{equation}}
\newcommand{\bea}{\begin{eqnarray}}
\newcommand{\eea}{\end{eqnarray}}

\begin{document}

\title{Comments on Noncommutative Particle Dynamics}

\author{Ciprian Acatrinei\thanks{On leave from: {\it Institute of
        Atomic Physics  -
        P.O. Box MG-6, 76900 Bucharest, Romania}; e-mail:
        acatrine@physics.uoc.gr.} \\
        Department of Physics, University of Crete, \\
        P.O. Box 2208, Heraklion, Greece}

\date{15 June, 2001}

\maketitle

\begin{abstract}
We discuss a version of Hamiltonian (2+1)-dimensional dynamics,
in which one allows nonvanishing Poisson brackets also between the coordinates,
and between the momenta.
The resulting equations of motion are not any more derivable from a Lagrangian.
However, taking a specific limit, in which the symplectic form becomes singular,
one can recover a first-order Lagrangian description.
This signals the dimensional reduction of the phase-space to half its initial
number of degrees of freedom.
We reach the same limit from another point of view, studying a particular form
of the Poisson brackets, which is singled out geometrically and easy to handle algebraically.
For comparison, a discussion of quantum mechanics with extended Heisenberg algebra is included.
The quantum theory constrains the antisymmetric matrix providing the algebra to the above
mentioned classically singular limit.
\end{abstract}

\section{Introduction}

Hamiltonian dynamics requires two basic types of data: 
the Hamiltonian $H$, 
and the symplectic two-form $\omega$ \cite{arnold}.
Whereas many types of Hamiltonians have been studied over the last two centuries,
the role of the symplectic form seems to have been often overlooked.
It is usually taken to be block-diagonalized from the start,
in the canonical form
$\omega=\sum_{i=1}^{n}dq_i\wedge dp_i$. 
The coordinates $q_i$ and the momenta $p_i$ represent the phase-space variables, 
denoted collectivelly  by $\{x_i\}$.

The aim of this note is to start a study of classical dynamics in presence of a more general,
but still constant,
canonical form
$\omega=\sum_{i,j=1}^{2n}dx_i\wedge dx_j$.
This corresponds to having extended Poisson brackets
$\{x_i,x_j\}=\Theta_{ij}$, where $\Theta$ is the inverse of $\omega$, 
$\Theta_{ij}=(\omega^{-1})_{ij}$.
A study of an $x-$dependent $\Theta(x)$ will be presented elsewhere. 

Of course, the constant matrix $\Theta_{ij}$ can be block-diagonalized
via non-canonical linear transformations, which mix the $q$'s and $p$'s.
One recovers then the usual canonical choice for $\omega$ and $\Theta$.
However, there are at least two reasons which might prevent one from doing so,
asking therefore for a study of dynamics with arbitrary  $\Theta$. 
First, although the canonical transformations of Hamiltonian dynamics
put $q$ and $p$ on equal footing, in the physical world there are always ways to distiguish
coordinates from momenta. Thus, to speak about physically relevant variables,
one may prefer to keep coordinates and momenta unmixed.
Second, the linear mixing of $q$'s and $p$'s required to block-diagonalize $\Theta$
transfers nonlinearity from the potential term of $H$ to the kinetic term.
Higher than quadratic terms in both momenta and coordinates 
being hard to deal with in Hamiltonian dynamics, 
one is pushed to test the dynamics in presence of the initial $\Theta$-matrix.
This is the point of view of this note. We will study, by elementary methods,
the dynamics of physical coordinates and momenta which do not 'commute' classically,
i.e. have nonvanishing Poisson brackets.
This is as close as one can come classically to noncomutative quantum mechanics,
which has generated much interest recently \cite{ncqm}.
The idea of  noncommutative quantum dynamics goes back to \cite{snyder}. 
For an interesting example of a 'noncommutative' classical system 
see for instance \cite{pt}.

We will work in two space-dimensions.
Since the Lagrangian formalism generates automatically canonical brackets,
our starting point will be the  Hamiltonian description of dynamics,
with extended Poisson brackets. We will write the extended Hamilton
equations of motion, and express them in terms of the coordinates $q_1$ and $q_2$.
Those equations are {\it not} derivable from a Lagrangian, except in some special cases.
However, if one takes the limit $\{q_i,q_j\}\{p_i,p_j\}\rightarrow 1$, 
in which case $\omega$ is singular,
one recovers a first-order Lagrangian description.
This signals a dimensional reduction taking place in phase-space, 
from four to two independent degrees of freedom.
We also notice that a minimally coupled gauge field loses its gauge invariance
property if $\{q_1,q_2\}\neq 0$.
Subsequently, we will approach the problem from the $\omega -$ block-diagonalization 
point of view. This is an involved problem in general. Still, a particular form of $\Theta_{ij}$
allows rapid progress, and one gets a simple diagonalizing transformation,
valid for a continuum af situations.
This will permit a somehow smooth transition to, 
and a different perspective on, the singular limit mentioned above.
For completeness, a study of the quantum mechanical extended Heisenbeg algebra,
$[x_i,x_j]=i \Theta_{ij}$, is presented in section 4. 
Quantum mechanics constrains the matrix $\Theta_{ij}$,
enforcing on it the classically singular limit $\det (\Theta)=0$.
Although such an observation does not seem to be new 
(for a more algebraic point of view, see \cite{corneliu}),
the proof we present seems quite illuminating, and might be helpful in connection
to recent work in noncommutative quantum mechanics \cite{ncqm}. 

\section{Equations of motion}

We start with the action
\be
S=\int dt \left ( \frac{1}{2}\omega_{ij}x_i x_j - H(x) \right ), \quad x_{1,2,3,4}=q_1,p_1,q_2,p_2,
\ee
which provides, upon independent variation of each $x_i$, the following equations of motion:
\be
\dot{x}_i=\{x_i,H\}=\Theta_{ij}\frac{\partial H}{\partial x_j}.
\ee
Above, $\Theta_{ij}=(\omega^{-1})_{ij}$, 
and $\{A,B\}$ is the extended Poisson bracket (PB) of $A$ and $B$.
In particular, $\{x_i,x_j\}=\Theta_{ij}$.
In order to render the expressions manageable, and for reasons detailed in sections  3 and 4,
we choose
\be
\Theta=
\left (
\begin{array}{rrrr}
0 & \theta & 1 & 0  \\
-\theta & 0 & 0 &1 \\
-1 &  0 & 0 & \sigma \\
0 & -1  & -\sigma &  0
\end{array} 
\right )
\quad \mbox{i.e.} \quad 
\omega=\frac{1}{1-\theta\sigma}
\left (
\begin{array}{rrrr}
0 & -\sigma & 1 & 0  \\
\sigma & 0 & 0 &1 \\
-1 &  0 & 0 &  -\theta\\
0 & -1  &   \theta & 0
\end{array} 
\right ).
\ee
Then the  nonvanishing Poisson brackets are
$\{q_i,p_j\}=\delta_{ij}$, $\{q_1,q_2\}=\theta$, $\{p_1,p_2\}=\sigma$, 
and the phase-space equations of motion become
\be
\dot{q}_i=
\frac{\partial H}{\partial p_i}+\theta \epsilon_{ij}\frac{\partial H}{\partial q_j},
\label{emq}
\ee
\be
\dot{p}_i=
-\frac{\partial H}{\partial q_i}+\sigma \epsilon_{ij}\frac{\partial H}{\partial p_j},
\label{emp}
\ee
with $\epsilon_{12}=-\epsilon_{21}=1$. If $\theta=\sigma=0$, 
(\ref{emq},\ref{emp}) are the usual Hamilton equations.

One remark is in order. By taking at the level of the equations of motion 
the  limit $\theta\sigma=1$, which renders $\Theta$ and $\omega$ singular,
one reduces by half the number of degrees of freedom having dynamics, since
\be
\dot{q}_1=-\theta \dot{p}_2 \quad \mbox{and} \quad \dot{q}_2=\theta \dot{p}_1.
\label{reduction}
\ee
We will have more to say about this later.

Let us express now the equations of motion in terms of $q_1$ and $q_2$.
Considering Hamiltonians of the form
\be
H=\frac{1}{2}(p_1^2+p_2^2)+V(q_1,q_2), \label{H1}
\ee 
one gets the coordinate equations of motion:
\be
\ddot{q}_i=-(1-\theta\sigma)\frac{\partial V}{\partial q_i}
+\sigma \epsilon_{ij} \dot{q}_j
+\theta \epsilon_{ij} \frac{d}{dt}\frac{\partial V}{\partial q_j}, \quad i=1,2.
\label{em1}
\ee
For $\theta\neq 0$, it is easy to see that equations (\ref{em1})
are not derivable from a Lagrangian, 
if the  potential $V$ is higher than quadratic in the coordinates.
The system does not appear to be dissipative, since a notion of conserved energy exists. 
For quadratic Hamiltonians, the effects of noncommutativity are trivially described;
they correspond to adding either constant forces, or magnetic fields, 
or harmonic oscillator potentials, into the Lagrangian for the commutative case.
We will not detail this here, but stress that
the classical effects of noncommutativity are truly relevant
only in the presence of nonlinear equations of motion.
The lack of a Lagrangian formulation reflects the difficulty in defining a suitable
Legendre transform if $\theta\neq 0$. 
In fact, if one would start from the Hamiltonian (\ref{H1}) 
and perform a usual Legendre transformation,
then use (\ref{emq}) to express $p_{1,2}$ in terms of $q_{1,2}$, 
one would get wrong equations of motion. 
The usual procedure works correctly only if $\theta =0$.
On the other hand, $\sigma$ is quite harmless; 
it plays the role of a constant magnetic field.

The RHS of (\ref{em1}) contains three kinds of terms.
The first, $-(1-\theta\sigma)\frac{\partial V}{\partial q_i}$, is just the usual
Newtonian force, apart from the $(1-\theta\sigma)$ factor.
The second term, $\sigma \epsilon_{ij} \dot{q}_j$, mimicks a magnetic field.
It is the third term which prevents the Lagrangian formalism from working.
However, if it is taken in isolation, it gives
\be
\ddot{q}_i=\theta \epsilon_{ij} \frac{d}{dt}\frac{\partial V}{\partial q_j}, 
\quad \mbox{i.e.} \quad 
\dot{q}_i=\theta \epsilon_{ij} \frac{\partial V}{\partial q_j}+c_i,
\quad i=1,2,
\label{third}
\ee
$c_i$ being two arbitrary constants.
(\ref{third}) allows a first-order Lagrangian formulation. 
Going for simplicity in a reference frame moving with velocity $\vec{v}=(c_1.c_2)$,
the equations $\dot{q}_i=\theta \epsilon_{ij} \frac{\partial V}{\partial q_j}$
are derivable from the Lagrangian
\be
L=\frac{1}{2}(\dot{q}_1q _2-q_1\dot{q}_2)-V(q_i).
\ee

Most important, the full equations (\ref{em1}) admit a first order Lagrangian description,
in the limit 
\be
\sigma\rightarrow\frac{1}{\theta}. \label{limit}
\ee
In this case, the usual Newtonian force term disappears completely
(this is kind of antipodal to usual Hamiltonian dynamics), and (\ref{em1}) becomes
\be
\dot{q}_i=\epsilon_{ij}(q_j/\theta+\theta\frac{\partial V}{\partial q_j})+C_i. \label{em2}
\ee
A first-order Lagrangian for (\ref{em2}) is:
\be
L=\frac{1}{2}(\dot{q}_1 q_2-\dot{q}_2 q_1)-\theta V(q_i)-
\frac{1}{2\theta}(q_1^2+q_2^2)+C_2q_1-C_1q_2. \label{L2}
\ee
The last two terms disappear in a suitably boosted reference frame.
Then the Lagrangian contains a term which is first order in time derivatives,
the usual potential $V$, and an additional two-dimensional harmonic oscillator potential.

The limit (\ref{limit}) reduces the number of degrees of freedom of the phase-space to half,
from four to two, cf. (\ref{reduction}). 
This is most clearly seen by noticing that (\ref{em2})  arises from the 
one-dimensional Hamiltonian
\be
H=\theta V(q,p)+\frac{1}{2\theta}(q^2+p^2)-C_2 q+C_1 p
\ee
after relabeling $q_1=q$, $q_2=p$. 
This is related to the dimensional reduction involved in the Peierls substitution 
\cite{peierls},
which is based on the noncommutativity of coordinates (see \cite{jackiw1}, 
which also refers to earlier work).
The connection is provided by the fact that 
in a 2D first-order system, the coordinates are canonically conjugate to each other 
\cite{jackiw2}.

Another simple type of Hamiltonian worth studying is
\be
H=\frac{1}{2}\sum_{i=1,2}(p_i-A_i(q_j))^2,
\ee
the gauge field $A_i$ being minimally coupled. Then
\be
\dot{q}_i=(p_l-A_l)(\delta_{il}-\theta\epsilon_{ij}\frac{\partial A_l}{\partial q_j}),
\label{a1}
\ee
\be
\dot{p}_i=(p_j-A_j)(\frac{\partial A_j}{\partial q_i}+\sigma\epsilon_{ij}).
\label{a2}
\ee
Assuming $\frac{\partial A_j}{\partial t}=0$, the pair (\ref{a1}) can be rewritten as
\be
p_i=A_i+\frac{1}{\Delta}\frac{d}{dt}(q_i+\theta\epsilon_{ij}A_j), \quad i=1,2 \label{b1},
\ee
where
$\Delta=1+\theta F_{12}+\theta^2\{A_1,A_2\}_{q_1q_2}$, with
$F_{12}=\partial_{1}A_2-\partial_{2}A_1$,
$\{A_1,A_2\}_{q_1q_2}=
\frac{\partial A_1}{\partial q_1}\frac{\partial A_2}{\partial q_2}-
\frac{\partial A_1}{\partial q_2}\frac{\partial A_2}{\partial q_1}$.
Using (\ref{b1}) in (\ref{a2}), and assuming 
$\frac{\partial A_1}{\partial q_1}=\frac{\partial A_2}{\partial q_2}=0$, 
one gets
\be
\ddot{q}_1=\left ( 1+\theta\frac{\partial A_1}{\partial q_2}\right )
\left [ 
-\dot{A}_1+\left (\frac{\partial A_2}{\partial q_1}+\sigma\right )
\frac{\dot{q}_2}{1-\theta\frac{\partial A_2}{\partial q_1}}
\right] \label{c1}
\ee
\be
\ddot{q}_2=\left (1+\theta\frac{\partial A_2}{\partial q_1}\right )
\left [
-\dot{A}_2+\left (\frac{\partial A_1}{\partial q_2}-\sigma\right )
\frac{\dot{q}_1}{1-\theta\frac{\partial A_1}{\partial q_2}}
\right ] \label{c2}
\ee
Let us consider the case of a constant magnetic field, 
$B=F_{12}=\partial_{1}A_2-\partial_{2}A_1$. 
This can be obtained in different gauges.
A striking feature of the equations (\ref{c1},\ref{c2}) is that they are {\it not}
gauge invariant, unless $\theta=0$.
For instance, in the symmetric gauge, $A_1=-q_2 B/2, A_2= q_1 B/2$, one has
\be
\ddot{q}_1=\dot{q}_2(\sigma+B+\theta B^2/4) \quad \quad
\ddot{q}_2=-\dot{q}_1(\sigma+B+\theta B^2/4),
\ee 
whereas in the gauge $A_1=0, A_2= q_1 B$ one gets
\be
\ddot{q}_1=\dot{q}_2\frac{(\sigma+B)}{(1+\theta B)} \quad \quad
\ddot{q}_2=-\dot{q}_1(\sigma+B)(1+\theta B),
\ee
which is not even derivable from a Lagrangian.
One sees again that $\sigma$ is inoffensive - it just adds to $B$ -
whereas $\theta$ even breaks gauge invariance!
A non-zero $\sigma$ is a good way to mimick a constant magnetic field,
without entering the difficulties caused by the lack of gauge invariance.

A reformulation of classical gauge theories appears necessary in this context. 
One possibility to restore gauge invariance is to introduce the star product
into the definition of $F_{ij}$ and of the gauge transformations of $A_i$.
This will not be discussed here.

\section{Another approach}

Let us return to the issue of block-diagonalization of $\Theta_{ij}$,
which may still be a useful thing to do for linear systems.
First, let us write the Poisson brackets in dimensionless form
\be
\{q_i/\sqrt{\theta},p_j\sqrt{\theta}\}=\delta_{ij}
\ee
\be
\{q_1/\sqrt{\theta},q_2/\sqrt{\theta}\}=1
\ee
\be
\{p_1\sqrt{\theta},p_2\sqrt{\theta}\}=\theta\sigma. \label{parameter}
\ee
We would like to discuss the meaning of the case $\theta\sigma=1$.
To do so, consider an hypothetical dimensionless Hamiltonian system
\be
\dot{X}_i=\bar{\Theta}_{ij}\frac{\partial H(X)}{\partial X_j},
\ee
with 
\be
\bar{\Theta}(t)=
\left (
\begin{array}{rrrr}
0 & t & 1 & 0  \\
t & 0 & 0 &1 \\
-1 &  0 & 0 & t \\
0 & -1  & -t &  0
\end{array} 
\right ).
\ee
To reach the limit $\theta\sigma=1$ we will take the singular limit $t\rightarrow 1$
at the end.
Since energy is conserved the flow (time derivative) of $X_i$ lies in the 
$(2n-1)-$dimensional hyperspace
orthogonal to $\nabla H=(\partial_1 H,\partial_2 H,\dots,\partial_{2n} H)^T$.
Thus, in one dimension, the situation is simple (the space orthogonal to 
$\nabla H=(\partial_q H,\partial_p H)^{T}$ being spanned by 
$(\partial_p H,-\partial_q H)^{T}$): 
conservation of energy determines the trajectory in phase space.
However, we are interested in two dimensional (2D) configuration spaces, 
hence in 4D phase spaces.										
The subspace of phase space in which motion is allowed will be 3-dimensional.
An orthogonal basis for this space is given by the vectors: 
\bea
A_2=
\left (
\begin{array}{rrrr}
0 & 0 & -1 & 0  \\
0 & 0 & 0 &-1 \\
1 &  0 & 0 & 0 \\
0 & 1  & 0 &  0
\end{array} 
\right )A_1, 
\quad   &
A_3=
\left (
\begin{array}{rrrr}
0 & -1 & 0 & 0  \\
1 & 0 & 0 &0 \\
0 &  0 & 0 & 1 \\
0 & 0  & -1 &  0
\end{array} 
\right )A_1, 
\quad  \\
A_4=
\left (
\begin{array}{rrrr}
0 & 0 & 0 & -1 \\
0 & 0 & 1 & 0 \\
0 & -1 & 0 & 0 \\
1 & 0 & 0 & 0
\end{array} 
\right )A_1,  &
\mbox{where}
\quad 
A_1=
\left (
\begin{array}{r}
\partial_{q_1} H \\ \partial_{p_1} H \\\partial_{q_2} H \\  \partial_{p_2} H
\end{array}
\right ).
\eea
The vector 
$A_1$
gives the direction along which motion is forbidden.
By abuse of language, we still denote the - now dimensionless - coordinates and momenta
by $q_i$, respectively $p_i$.
At a given moment, 
the system flows along a particular linear combination of $A_2$, $A_3$ and $A_4$.
The existence of a flow along  $A_2$  can be seen  as due to 
nonvanishing (and equal) Poisson brackets
between $q_1$ and $p_1$, and between $q_2$ and $p_2$. The canonical formulation,
in which $\omega=dq_1 \wedge dp_1+dq_2\wedge dp_2$, corresponds to that case.
On the other hand,
a flow along  $A_3$  corresponds to nonvanishing (and equal) Poisson brackets
between $q_1$ and $q_2$, and between $p_1$ and $p_2$,
$\{q_1,q_2\}=\{p_1,p_2\}=t\neq 0$.
The existence of a flow along  $A_4$  would correspond to  Poisson brackets
between $q_1$ and $p_2$, and between $q_2$ and $p_1$. We neglect this possibility,
which is anyway prohibited by quantum mechanics, as we will show in the next section.
Having $\{q_1,q_2\}=t\neq s=\{p_1,p_2\}$ means allowing flows in the whole
3D space available.
Thus the case $t=s$ restricts motion to a 2D subspace, spanned by $A_{2,3}$,
and is expected to be simpler to study, as we will see.
Taking then $t\rightarrow 1$ will allow 
further insight in the  $\theta\sigma=1$ limit.

To begin, we diagonalize the general matrix 
\be
\Theta(t,s)=
\left (
\begin{array}{rrrr}
0 & t & 1 & 0  \\
t & 0 & 0 &1 \\
-1 &  0 & 0 & s \\
0 & -1  & -s &  0
\end{array} 
\right ).
\ee
Its eigenvalues are
\be
\lambda_{1,2}^2=-\left (
1+\frac{s^2+t^2}{2}
\right )
\stackrel{+}{-}
\sqrt{\left(1+\frac{s^2+t^2}{2}\right)^2-(1-ts)^2}.
\ee
They have a simple form if $t=s$:
\be
\lambda_1=-i(1+t),\quad \lambda_2=i(1+t),
\quad \lambda_3=-i(1-t), \quad \lambda_4=i(1-t),
\ee
and their corresponding orthonormal 
eigenvectors are
\be
v_1=\frac{1}{2}
\left (
\begin{array}{r}
-1 \\ i \\ i \\ 1
\end{array}
\right ) 
\quad  
v_2=\frac{1}{2}
\left (
\begin{array}{r}
-1 \\ -i \\ -i \\ 1
\end{array}
\right )  
\quad  
v_3=\frac{1}{2}
\left (
\begin{array}{r}
1 \\ -i \\ i \\ 1
\end{array}
\right )  
\quad 
v_4=\frac{1}{2}
\left (
\begin{array}{r}
1 \\ i \\ -i \\ 1
\end{array}
\right ). 
\quad
\ee
Remarkably, these eigenvectors, and the diagonalization transformation they generate,
are independent of $t$. 
This will allow the otherwise singular limit $t\rightarrow 1$ 
to be taken at the level of the eigenvectors.
The coordinates which block-diagonalize $\Theta$, 
namely $Q_1,P_1,Q_2,\mbox{ and } P_2$, are given by
\be
Q_1=\frac{(p_2-q_1)}{\sqrt{2}}, \quad  P_1=\frac{(-q_2-p_1)}{\sqrt{2}}, \quad
Q_2=\frac{(q_1+p_2)}{\sqrt{2}}, \quad  P_2=\frac{(q_2-p_1)}{\sqrt{2}};
\ee
they have only two nonvanishing Poisson brackets, 
\be
\{Q_1,P_1\}=1+t   \quad \quad   \{Q_2,P_2\}=t-1.
\ee
In the $t\rightarrow 1$ limit (corresponding to $\theta\sigma=1$ in (\ref{parameter}))
the influence of $H$ on the coordinates $Q_2$ and $P_2$ decreases,
due to the decrease in magnitude of their Poisson brackets.
At $t=1$, $Q_2$ and $P_2$ become simple parameters, all the dynamics taking place
only in the $(Q_1,P_1)-$space. This is the dimensional reduction formerly advertised.

We also notice that 
$q_1^2+q_2^2+p_1^2+p_2^2=Q_1^2+Q_2^2+P_1^2+P_2^2$;
the diagonalization transformation belongs to the $SO(4)-$symmetry group of phase space.
Thus, noncommutativity is irelevant for a 2D harmonic oscillator.
Of course this is valid, with trivial additional specifications from case to case,
for all quadratic systems.
Noncommutativity starts playing effectively its role, 
and renders problems even more difficult  to handle,
once nonlinearities are present.

\section{Noncommutative quantum mechanics}

We saw that in classical mechanics the singulat limit 
$\sigma=1/\theta$ 
- which can be taken at the level of the equations of motion -
has two special features.
It admits a first-order Lagrangian description (related to dimansional reduction),
and geometrically it means that one allows flows only in a 2D part of the 3D
subspace of phase-space available for motion. Algabraically, it is easier to handle
than the general set-up.

We are going to show now that this specific limit,  $\sigma=1/\theta$,
is {\it enforced } by quantum mechanics.
Although taking this limit appears to be common lore
(one discussion appeared in \cite{corneliu}),
the proof to be presented here is new, and might be of interest in connection to
the recent work in noncommutative quantum mechanics \cite{ncqm}. 
It is also hoped it will help to understand the apparent discrepancy
between the  quantum ($\sigma=1/\theta$) and classical ($\sigma\neq1/\theta$) regimes. 

Let us consider the extended Heisenberg algebra (with $\hbar=1$, for simplicity)
\be
[\hat{q}_i,\hat{p}_j]=i\delta_{ij} 
\quad [\hat{q}_1,\hat{q}\hat{q}_2]=i\theta 
\quad  [\hat{p}_1,\hat{p}_2]=i\sigma \quad
[p_i,\epsilon_{ij}q_j]=0. \label{cr}
\ee
We start by assuming that $q_1$ and $p_2$ are independent variables
(an assumption we will have to reconsider soon).
Then one can speak about the wave function $\psi(q_1,p_2)$, 
since $[\hat{q}_1,\hat{p}_2]=0$.
Employing now the commutation relations (\ref{cr}), one obtains,
depending on whether one applies $\hat{p}_1$ to the first or the second variable
in $\psi(q_1,p_2)$,
\be
\hat{p}_1\psi(q_1,p_2)=
\frac{1}{i}\partial_{q_1}\psi(q_1,p_2)=
-\frac{\sigma}{i}\partial_{p_2}\psi(q_1,p_2). 
\label{AA}
\ee
A similar reasoning applies for $\hat{q}_2$:
\be
\hat{q}_2\psi(q_1,p_2)=
-\frac{1}{i}\partial_{p_2}\psi(q_1,p_2)=
\frac{\theta}{i}\partial_{q_1}\psi(q_1,p_2). 
\label{BB}
\ee
Consistency of (\ref{AA}) and (\ref{BB}) obviously requires
\be
\theta\sigma=1. \label{CC}
\ee
Once (\ref{CC}) is assumed,
it is easy to show that $q_i+\theta^{-1}\epsilon_{ij}p_j=0$, up to the unity operator,
since these expressions commute with everything.
Thus, $q_1$ and $p_2$  are {\it not } independent variables; 
the wave function
feels the dimensional reduction, and it becomes either $\psi(q_1)$, or $\psi(p_2)$.
Alternatively, if one starts with the correct assumption $\hat{p}_2\sim\hat{q}_1$,
relation (\ref{CC}) is trivially obtained through manipulations of (\ref{cr}).

One may question the above argument, as it involves unbounded operators
(the same applies to \cite{corneliu}). 
To counter such an objection,
we provide an argument using the following bounded operators 
\be
\hat{U}(\alpha)=e^{i\alpha\hat{q}_1} \quad
\hat{V}(\beta)=e^{i\beta\hat{p}_1} \quad
\hat{W}(\gamma)=e^{i\gamma\hat{q}_2} \quad
\hat{Z}(\delta)=e^{i\delta\hat{p}_2}. 
\ee
Assuming again that $\hat{q}_1$ and $\hat{p}_2$ are independent,
$\{ |q_1,p_2>\}$ is a complete set of eigenvectors.
Using (\ref{cr}), one can show, after some manipulations, that
\be
e^{i\beta\hat{p}_1} |q_1,p_2>=|q_1+\beta,p_2>=|q_1,p_2-\beta \sigma>  
\label{AAA}
\ee
\be
e^{i\gamma\hat{q}_2}|q_1,p_2>=|q_1+\gamma\theta,p_2>=|q_1,p_2-\gamma>. 
\label{BBB}
\ee
The above relations are exact up to normalization factors. They constrain again
$\theta$ and $\sigma$ by (\ref{CC}).
The dimensional reduction can be nicely seen at work in the $(q_1,p_2)$ plane, 
which labels the eigenvectors $|q_1,p_2>$. 
The relation (\ref{AAA}) identifies the  points 
$(q_1+\beta,p_2)$ and $(q_1,p_2-\beta \sigma)$ in this plane,
whereas (\ref{BBB}) identifies $(q_1+\gamma\theta,p_2)$ and $(q_1,p_2-\gamma)$.
Let us denote the vectors between these two pairs of points 
by $\vec{e}_1$ and $\vec{e}_2$.
If $\sigma\theta =1$, $\vec{e}_1$ and $\vec{e}_2$ are parallel.
Now, since $\beta$ and $\gamma$ can take any real value,
all the points along any line parallel to $\vec{e}_1$ (or $\vec{e}_2$) are identified.
Consequently, the plane $(q_1,p_2)$ reduces to a line, which is orthogonal to $\vec{e}_1$. 
The slope of this line is $-\frac{1}{\theta}$, thus the line satisfies the equation
$p_2=\frac{1}{\theta}q_1+cst$. 
The linear dependence of the eigenvalues transfers to the operators: 
$\hat{p}_2=-\frac{1}{\theta}\hat{q}_1$. A similar argument shows that 
$\hat{p}_1=-\frac{1}{\theta}\hat{q}_2$. We demonstrated the dimensional reduction once more.
Again, (\ref{CC}) follows immediately once the previous relations are assumed.
If $\theta\sigma\neq 1$, then $\vec{e}_1$ and $\vec{e}_2$ are not parallel:
the whole $(q_1,p_2)$ plane would collapse to a single point.

Two remarks are in order. 
We assumed that $[p_i,\epsilon_{ij}q_j]=0$. If one allows
nonvanishing commutators $[\hat{q}_1,\hat{p}_2]$ and  $[\hat{q}_2,\hat{p}_1]$ too, 
arguments similar to the above ones 
show that all the operators of the algebra (\ref{cr}) 
are proportional to each other - a contradiction. 
This is why we did not consider 
such nonvanishing Poisson brackets in the previous two sections. 

One may escape the above conclusions by assuming
that the operators $\hat{q}_i$ and $\hat{p}_i$ act in a direct product of Hilbert spaces,
with one pair of commutation relations  ($[\hat{q}_i,\hat{p}_i]$ say) enforced in one space,
and the other pair ($[\hat{q}_1,\hat{q}_2]$ and $[\hat{p}_1,\hat{p}_2]$) working in the second space. 
The above arguments do not apply to this case. 

\section{Conclusion}

We have studied some particularities of 'noncommutative' classical dynamics,
as represented by Hamilton's  equations in the presence of a non-diagonal symplectic form.

For quadratic systems, the effects of noncommutativity are reproduced
by commutative theories with additional, at most quadratic,  terms in the action.
For nonlinear systems, noncommutativity prevents the equations of motion
from being put into Lagrangian form, 
and renders them even more difficult to solve than in the commutative case.

An exception appears when the symplectic form is singular.
Then, only half of the degrees of freedom are dynamical,
as the existence of a first-order Lagrangian formulation also illustrates.


In contrast, quantum mechanics enforces the above dimensional reduction, 
which appeared in classical mechanics only in a particular limit.
This different behaviour deserves a better study,
as it is a basic example in which the correspondence principle does not seem to apply:
quantum mechanics imposes additional constraints on parameters which
appear to be free in the classical world.

\subsection*{Acknowledgements}
I am grateful to Gianpiero Mangano for very useful discussions and help
during the initial stages of this work.
I thank the HEP theory group of the University of Crete for kind hospitality,
and E. Kiritsis, C. Sochichiu and  T. N. Tomaras for useful comments.
This work was supported by funds of the Italian Ministry for Education and Research.

\begin{thebibliography}{1}

\bibitem{arnold}
V. I. Arnold,
\newblock {\it Mathematical Methods of Classical Mechanics}, Springer 1989.
\newblock 

\bibitem{ncqm}
An incomplete list of references is:

G. Mangano, {\it J. Math. Phys.} {\bf39}, 2584 (1998), gr-qc/9705040,  

C. Duval, P. A. Horvathy,
{\it Phys. Lett.} {\bf B479}  284 (2000), hep-th/0002233,

L. Mezincescu, hep-th/0007046,

V.P. Nair, A.P. Polychronakos, hep-th/0011172, 
 
B. Morariu, A. P. Polychronakos, hep-th/0102157,

J. Gamboa, M. Loewe, J.C. Rojas, hep-th/0010220.

\bibitem{snyder} H.S. Snyder, {\it Phys. Rev.} {\bf 71}, 38 (1947).

\bibitem{pt} N. Papanicolaou and T.N. Tomaras, {\it Nucl. Phys.} {\bf B360}, 425 (1991).

\bibitem{corneliu}
C. Sochichiu, hep-th/0010149.

\bibitem{peierls}
R. Peierls, {\it Z. Phys.} {\bf 80} 763 (1933).

\bibitem{jackiw1}
G. Dunne , R. Jackiw
{\it Nucl. Phys. Proc. Suppl.} {\bf 33C}, 114 (1993), 
hep-th/9204057.

\bibitem{jackiw2}
 

L. Faddeev and R. Jackiw, 
{\it Phys. Rev. Lett.} {\bf 60}, 1692 (1988), 

R. Jackiw, hep-th/9306075.

\end{thebibliography}

\end{document}